# Post-human interaction design, yes, but cautiously

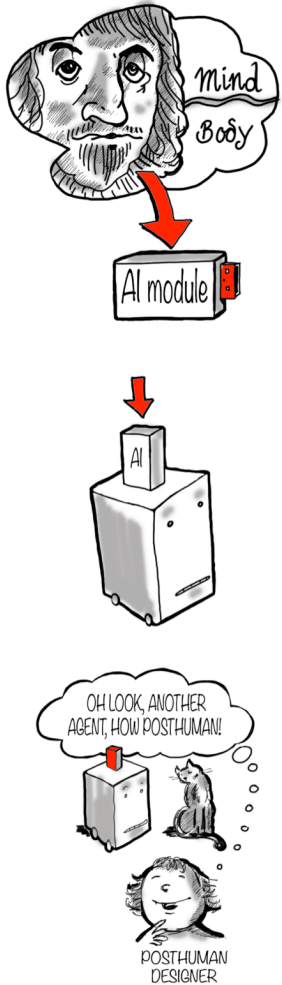

Figure 1. Cartesian logic, embedded in AI technology. implicitly imported into post-human smart objects Images © Jelle van Dijk

**Jelle van Dijk**
University of Twente
P.O.Box 217, 7500 AE
Enschede
Jelle.vandijk@utwente.nl

## Abstract
Post-human design runs the risk of obscuring the fact that AI technology actually imports a Cartesian humanist logic, which subsequently influences how we design and conceive of so-called 'smart' or 'intelligent' objects. This leads to unwanted metaphorical attributions of human qualities to 'smart objects'. Instead, starting from an 'embodied sensemaking' perspective, designers should demand of engineers to radically transform the very structure of AI technology, in order to truly support critical posthuman values of collectivity, relationality and community building.

**CSS Concepts**
• Human-centered computing~Interaction Design theory, concepts and paradigms

## Introduction
It is a good thing that Interaction Design has embraced the post-human. For ages, humans have seen themselves as the epicenter of the world. Given the global challenges today, it seems long due we reconsider our proper place in the complex, interconnected eco-system that is planet earth. In design, this has lead to the recognition that 'human-centred design' is in a way problematic, as it would suggest that design is for the benefit of humans only, in disregard of the interests of non-humans such as animals, or even rivers.

But when design 'beyond the human' goes together with Artificial Intelligence (AI) there is a risk. The risk is that post-human design will not prevent, and may even obscure, the fact that AI technology itself imports a traditional, humanist form of logic, which subsequently influences how we design and conceive of so-called 'smart' or 'intelligent' objects. In this provocation I elaborate this risk and offer an alternative frame, based on the *enactivist* notion of *embodied sense-making*.

Starting with the view that human beings are embodied sense-makers, I argue that the 'smart objects' we design are not themselves 'sense-makers', they are at best *part of* our extended living bodies, through which *we* make sense. Smart objects remain *things that humans think with'*, not 'thinking things'. An enactive embodied view allows us to throw away a lot of unwanted Humanist bathwater, while retaining a human – that is, an embodied, situated and ecologically sustainable human - baby.

## The rise of the smart object

Smart algorithms used today were developed in the cognitive sciences in the 1980s and 1990s, yet it is only recently that we see their massive deployment in real-world contexts. This is largely due to the explosion of available data. Thus we see a wealth of research on smart or intelligent objects, products, and systems, and a strong interest for AI in interaction design research. Several design-researchers in the area of smart, consumer-type artifacts now explicitly endorse a post-human frame to re-conceptualize the human-technology relations involved [1]. Yet somehow, this has led to the idea that the 'smart objects' that will be inhabiting our lives should be granted a status of autonomous sense-makers, such as we read in [2]:

"the question then is how to design intuitive collaborations between humans and non-humans. … Considering intelligent products as agents acknowledges that they sense, respond, and cooperate in human activity in an *autonomous manner*."

Or as Ron Wakkary puts it: "By posthuman, I mean thinking about the world as if humans share center stage with non-humans ….things are made of matter that is vibrant and agentic such that they appear to have "a life of their own" and so relate to us more like companions than tools."[1]

My provocation is this: the idea that objects truly have a life of their own, that they 'autonomously' sense and cooperate with humans as companions, is false. It can be useful in a metaphorical sense, but note it is essentially a *humanist* conception of humans (as thinking agents, socially interacting with other agents, responding intelligently to the environment), applied to an artifact. The metaphor can also have negative ethical consequences, to which I turn later. Most importantly however, this metaphor is rooted deep into the very structure of AI itself, and instead of building on it, design should instead concern itself with getting rid of it once and for all.

## AI is Humanistic

Imagine a hypothetical artificial neural network (ANN). Suppose the network takes as input pattern pixel values of a camera and, after training, reliably outputs 1 when the camera image shows a person and 2 when it shows a cat. Imagine the AI module is used as a design part in a smart object moving around in a household. This little critter is designed to have certain appropriate responses cat, and others that apply to its human household members. Taking a post-human scenario, we can now research and analyse the activities in the household from the perspectives of the humans, the cat, and from the view of the artifact.

---

[1] https://uwaterloo.ca/games-institute/events/guest-lecture-things-we-could-design-more-human-centred, retr. 13/3/20.

This sounds reasonable at first sight. However, note that the output of the neural network (cat, human) is already a **human** judgment, implemented by the designer, a human being, for whom cats and humans are sensible categories. We tend to say the network 'learned the distinction by itself'. But what really happens is that engineers implemented *their* categories of humans and cats using as a functional mechanism the property of ANNs being able to settle iteratively on statistical regularities in large sets of data. The goal states that the network trains on are human categories.

Furthermore, the distinction (cat, human) is a human ***judgment***. That is to say: the software in the system acts as a *cognitive* system: it models the outside world internally, and reaches a cognitive *judgment*, before it acts. We may talk about the critter as an 'embodied' agent, but in reality it is a physical machine, on the one hand, with a detached 'mind' in the form of its algorithms, on the other. If these two work well together, this is due to skilled craftsmanship of designers only. Any meaning we attribute to its behavior is grounded not 'in *its* being' but in *our, human* being: *we* humans, make sense of what the artifact is doing and *we* make sure the artifact is doing something that makes sense.

Consider even the sensor. Sensors are material artifacts, crafted by humans, that transform physical qualities to a digital signal, say, a level between 0 and 5 Volt. A sensor being active means *representing the world:* the exterior environment is *measured* and re-presented as a digital 'state'. This transformation turns everything that happens after into a process modeled on the Cartesian mind: the inner 'cognitive' system sensing what is 'going on outside', eventually, to control it. Moreover, the sensor is designed to sample some qualities and not others, with a preset resolution. Such engineering choices are ultimately grounded in human understanding of the world, not in any intrinsic understanding 'by the artifact'. Interactive technology, especially when equipped with AI, does not generate a new 'being' living together with humans: it implements a rather traditional, Cartesian model of human cognition, represented in a crude mechanical simulacrum (See Figure 1).

**Embodied sense-making**
In biological organisms, what we call 'sensors' and 'neural processing' are integrated aspects of living bodies, gradually evolved in interaction with ongoing, self-sustaining sense-making activity [3]. Nobody designed our sensors, and they did not come with prefab meaning. Our 'sensors' are so deeply interwoven with our action system that the distinction between 'sensor' and 'actuator' falls short: we are living, sense-making bodies, sensor and 'motor' at the same time. From an embodied sense-making perspective and in line with mediation theory [4] technologies are incorporated in our human sense-making practices, and our sense-making bodies become extended bodies. AI technology changes nothing to this basic scheme. However, AI itself is modeled on quite a different worldview: a rationalist, Cartesian, that is essentially Humanist conception of mind as standing over and against the world. Smart objects, even in post-human design are still tainted by these Humanist roots.

**Discussion**
It may feel natural and harmless to design objects as if they can hear, see, feel, and interact with us socially. But the illusion created by these metaphors may actually be harmful. Thus, we may be tempted to see Roomba the vacuumcleaner as a rudimentary "Sophia" – the famous humanoid, i.e. as a thinking social agent, some-one (not some-thing), able to 'take a perspective' and engage in social interactions with 'other' household members (most notably, of course, the cat). But we should better understand Roomba as not an independent household member, but rather as a sensorial extension of human embodied sense-making. Mind you, not of the household members themselves, but of the data scientists in the commercial companies that wish to observe and gather data from consumers. Roomba is not R2D2, but James Bond's spy-camera, disguised as a ballpoint pen.

Embodied sense-making radically changes the conception of ourselves as rational minds with physical bodies. Yet it also rejects the idea of interactive technologies as 'also' embodied sense-makers. This is not a move back into traditional 'human centered' design. It aligns with Braidotti's 'critical post-humanism', when she writes: "Posthuman subjectivity expresses an embodied and embedded and hence partial form of accountability, based on a strong sense of collectivity, relationality and hence community building."[5]

To use technologies for such community building is precisely what makes us *more* human, in contrast to the Humanist project that saw technology as a way to for Western colonists to eventually dominate the world. But this is far removed from giving any special status to smart objects. It is difficult to get rid of this fantasy of intelligence in machines, especially when designers take AI as a ready-made 'module', along with all narratives, to embed in their artifacts. It gets even worse when designers keep using metaphors for artifacts as 'observing', 'understanding', 'desiring', and so on, which are all essentially Cartesian, humanist, descriptions of human being.

If used with care, agent metaphors can be very useful. Giaccardi developed an elegant ethnography of the thing: "movie clips [shot] from the perspective of our three objects, help[ed] explore what these objects 'experience'". [6] To think 'from' the object's perspective, metaphorically, may help *create human value* (in this case, people with dementia). However, interaction designers can actually go much further. They should start concerning themselves with criticizing *the actual technical structure of AI.* Based on their holistic, skilled designerly intuitions, and guided by a deep empathic involvement with real-world human practices, they should not just design metaphorical interfaces, but instead demand of AI engineers that their Cartesian mechanics need to be replaced by a form of interactive processing that truly supports the posthuman collectivity, relationality and community building advocated by Braidotti.

**Conclusion**
The position advocated here could be characterized as a cautious, critical form of post-humanism. It can also be taken as a 'Humanism 2.0' – arguing that our design concerns are still, in the end *human* concerns, even if being human means a fully embodied and situated in - and striving to live in harmony with - the larger eco-system of animals and matter that makes up our planet. It may take a change not just in appearance but in deep structure of 'smart' objects to be able to fully cater such values. As a start, this paper hopefully helps design-researchers get clear on what is currently at stake in the design of interactive objects.